# Thermal Expansion in 3d-Metal Prussian Blue Analogs – A Survey Study

Sourav Adak[a,b], Luke L. Daemen[b], Monika Hartl[b], Darrick Williams[c], Jennifer Summerhill[d] and Heinz Nakotte[a]

[a]Department of Physics, New Mexico State University, Las Cruces NM 88003, USA
[b]Los Alamos Neutron Science Center, Los Alamos National Laboratory, Los Alamos NM 87545, USA
[c]Center for Integrated Nanotechnologies, Los Alamos National Laboratory, Los Alamos NM 87545, USA
[d]Department of Chemistry and Biochemistry, New Mexico State University, Las Cruces NM 88003, USA

**Abstract**

We present a comprehensive study of the structural properties and the thermal expansion behavior of 17 different Prussian Blue Analogs (PBAs) with compositions $M^{II}_3[(M')^{III}(CN)_6]_2 \cdot nH_2O$ and $M^{II}_2[Fe^{II}(CN)_6] \cdot nH_2O$, where $M^{II}$ = Mn, Fe, Co, Ni, Cu and Zn, $(M')^{III}$ = Co, Fe and n is the number of water molecules, which range from 5 to 18 for these compounds. The PBAs were synthesized via standard chemical precipitation methods, and temperature-dependent X-ray diffraction studies were performed in the temperature range between -150°C (123 K) and room-temperature. The vast majority of the studied PBAs were found to crystallize in cubic structures of space groups $Fm\bar{3}m$, $F\bar{4}3m$ and $Pm\bar{3}m$. The temperature dependence of the lattice parameters was taken to compute an average coefficient of linear thermal expansion in the studied temperature range. Of the 17 compounds, 9 display negative values for the average coefficient of linear thermal expansion, which can be as large as 39.7 x 10$^{-6}$ K$^{-1}$ for $Co_3[Co(CN)_6]_2 \cdot 12H_2O$. All of the $M^{II}_3[Co^{III}(CN)_6]_2 \cdot nH_2O$ compounds show negative thermal expansion behavior, which correlates with the Irving-Williams series for metal complex stability. The thermal expansion behavior for the PBAs of the $M^{II}_3[Fe^{III}(CN)_6]_2 \cdot nH_2O$ family are found to switch between positive (for M = Mn, Co, Ni) and negative (M = Cu, Zn) behavior, depending on the choice of the metal cation (M). On the other hand, all of the $M^{II}_2[Fe^{II}(CN)_6] \cdot nH_2O$ compounds show positive thermal expansion behavior.

## 1. Introduction

When heated, most materials expand by virtue of the inherent anharmonicity of the vibrations of its chemical bonds. The average distance between bonded pairs of atoms increases with temperature, and this increase usually results in an expansion of the material at the macroscopic scale [1]. A quantitative measure of the expansion behavior is provided by the coefficient of linear thermal expansion

$$\alpha = \frac{\Delta L}{L_0 \Delta T}$$

where $\Delta L = L - L_0$, $\Delta T = T - T_0$, $L$ is the lattice parameter at room temperature and $L_0$ is the lattice parameter at another reference temperature. For most materials, the coefficient of linear thermal expansion, α, is usually within the range $10^{-7}$-$10^{-5}$ K$^{-1}$. However, there are a few materials that actually shrink when heated. This effect is commonly referred to as Negative Thermal Expansion (NTE).

NTE materials are of considerable interest for a variety of technological applications, since they allow designing devices and support structures with precisely tailored coefficients of thermal expansion. It is possible to achieve this goal using a NTE material incorporated into composites, as suitable NTE materials can compensate for the more usual positive thermal expansion (PTE) behavior of other materials. For example, telescope mirror blanks require essentially "zero" thermal expansion (a few parts per million in the dimensional changes) over the range of temperatures found at the telescope location. On the other hand, the design of a 'zero-expansion' material over an extended temperature range has been a daunting task for many other evolving technologies, such as aerospace applications, gas turbine engines, electronic circuit boards, household items (such as cookware), and dentistry. The increased demand for materials with particular thermal expansion properties, paired with scientific curiosity, has geared the attention of the NTE scientific community toward identifying the mechanism(s) responsible for this unusual behavior in different classes of NTE materials.

Although NTE behavior is fairly uncommon, it has been reported in several classes of materials, such as alloys [2], intermetallic compounds [3], oxides [4-9], zeolites [10] and framework structures [11-18]. Belonging to the framework structures, NTE behavior has been reported for several compounds of Prussian Blue Analog (PBA) structural families [15-18], and we will briefly summarize some of those findings next. Chapman *et al.* [15] studied the NTE of PBAs of the type *MPt(CN)₆* with *M = Mn, Fe, Co, Ni, Cu, Zn, Cd* in the temperature range from 100 to 400 K, and they observed NTE behavior for all of these compounds with coefficients for $\alpha$ ranging from about $-1 \times 10^{-6}$ K$^{-1}$ (*M =Ni* ) to about $-10 \times 10^{-6}$ K$^{-1}$ (*M =Cd*). Margadonna *et al.* [16] reported negative or near-zero thermal expansion behavior for the cubic *Fe[Co(CN)₆]* with $\alpha = -1.5 \times 10^{-6}$ K$^{-1}$ in the temperature range from 4.2 to 300 K. These authors argued that NTE in this PBA can be attributed to the rigid unit modes vibrations of the stiff octahedral units. Goodwin *et al.* [17] studied NTE behavior in PBAs of type *M$^{II}$Pt$^{IV}$(CN)₆.nH₂O* with *M = Zn, Cd* and *0 ≤ n ≤ 2*. They found NTE in *ZnPt(CN)₆.0H₂O* ($\alpha = -3.38 \times 10^{-6}$ K$^{-1}$), *CdPt(CN)₆.2H₂O* ($\alpha = -7.31 \times 10^{-6}$ K$^{-1}$), and *CdPt(CN)₆.0H₂O* ($\alpha = -6.69 \times 10^{-6}$ K$^{-1}$). In a recent paper, Matsuda *et al.* [18] reported positive and negative thermal responses of a series of the PBAs of type *(Cs,Rb)$_x$M$^{II}$[Fe(CN)₆]$_y$.nH₂O* with *M = Ni, Cu, Zn, Cd.* As a consequence of such studies, the scientific community adopted a "common knowledge" that PBAs are prone candidates for NTE. However, considering the large number and variety of possible PBAs, only relatively few have been studied in detail, and a systematic and comprehensive investigation of the linear thermal expansion coefficients in PBAs has not been performed yet. Moreover, as evidenced by Masuda's study [18], not all PBAs will exhibit NTE behavior. Because it is possible to systematically vary ion size and charge in PBAs, they present an attractive playground to study systematically the occurrence of NTE in that family of materials, and thus explore its possible correlations with the electronic and crystal structures.

In this paper, we present a systematic investigation of the thermal expansion behavior for three families of Prussian Blue Analogs, namely

(1) the *hexacyanocobaltates(III)* of composition $M^{II}_3[Co^{III}(CN)_6]_2.nH_2O$,

(2) the *hexacyanoferrates(III)* of composition $M^{II}_3[Fe^{III}(CN)_6]_2.nH_2O$, and

(3) the *hexacyanoferrates(II)* of composition $M^{II}_2[Fe^{II}(CN)_6].nH_2O$.

In the chemical formulas above, *M* refers to a first-row (*3d*) transition metal (i.e. *M = Mn, Fe, Co, Ni, Cu* or *Zn*), *n* refers to the number of water molecules per formula unit, and the superscripts *II* or *III* indicate the expected oxidation states of the respective metal ions. A general structural feature of PBAs is that they consist of two types of metal centered octahedral structural units linked through cyanide chains.

The main goal of this paper is to determine and tabulate the thermal expansion coefficients of the above PBAs, rather than a detailed study and a discussion of the underlying mechanisms of the thermal expansion behavior of particular compounds or family of PBAs.

## 2. Experimental methods

### 2.1 Sample Synthesis

The synthesis of PBAs through standard chemical precipitation methods is fairly straightforward. However, product purification requires a careful and extensive washing procedure. We used ACS quality reagents without further purification. The appropriate metal nitrates or chlorides (of *Mn, Fe, Co, Ni, Cu* and *Zn*) were used for the synthesis of the *hexacyanocobaltates(III)*, the *hexacyanoferrates(III)*, and the *hexacyanoferrates(II)*. We were able to synthesize 17 of the possible 18 different PBAs as stable single-phase materials.

For the synthesis of *hexacyanoferrates(III)*, two separate solutions were prepared: a) one where *37.5 mmol* of metal nitrate was dissolved into *50 ml* water, and b) a second one where *25 mmol* of potassium ferricyanide *($K_3[Fe(CN)_6]$)* was dissolved into *250 ml* water. The first solution was poured into the second one under vigorous stirring. The solid precipitate was filtered out, washed multiple times with large amounts of water, and dried in air overnight at room temperature. Finally, grinding with a mortar and pestle produced fine powders suitable for X-ray diffraction studies. *Hexacyanocobaltates(III)* were similarly synthesized starting with *$K_3[Co(CN)_6]$*. For the synthesis of *hexacyanoferrates(II)*, we dissolved *15 mmol* (*6.34 g*) of *$K_4[Fe(CN)_6] \cdot 3H_2O$* in *100 ml* water and *30 mmol* of the relevant metal nitrate in *50 ml* of water and then followed the same steps as above. It is important to note that potassium is not usually incorporated in the framework of the PBAs and can be removed by repeated washing. None of our PBAs was exposed to any additional heat treatment prior to the experimental studies, leaving us with the fully hydrated PBA compounds.

### 2.2 Sample Characterization

After synthesis, washing and drying, all of our PBAs were characterized using standard characterizations techniques, namely X-ray diffraction (XRD), X-ray fluorescence (XRF), Fourier-transform infrared (FTIR) spectroscopy, and thermogravimetric analysis (TGA).

*2.2.1 Crystal structure characterization via X-ray powder diffraction (XRD)*

To characterize the crystal structures of the synthesized samples, room temperature XRD patterns were collected using a Rigaku Ultima III X-ray diffractometer with monochromatic CuK$_\alpha$ (*λ = 1.54 Å*) radiation in the *2θ* range 10–90º with a scan speed 2º/min and step size 0.04º. The diffractometer was operated in Bragg-Brentano geometry. For the XRD measurements, powder samples were loaded on a flat copper plate. The diffraction patterns were used to determine the phase purity and the crystal structure of each of our PBAs. Using the above synthesis approach, most of our PBAs synthesize as single phase materials, except for *Fe$_2$[Fe(CN)$_6$].nH$_2$O,* which was found to be unstable. Using the Rietveld refinement program package, General Structure Analysis System (GSAS) [19], we established that the studied PBAs synthesize in various crystal structures shown in Figure 1 with lattice and structural parameters given in Tables 1 and 2. Most of our PBAs crystallize in the cubic space groups $Fm\bar{3}m$, $F\bar{4}3m$ or $Pm\bar{3}m$ (Figures 1a-c). *Mn$_2$[Fe(CN)$_6$].9H$_2$O* and *Zn$_2$[Fe(CN)$_6$].5H$_2$O*, on the other hand, crystallize in the monoclinic $P2_1/n$ and trigonal $P\bar{3}$ space groups (not shown in Figure 1), respectively, which are distortions of the cubic structures. All structures invariably consist of two types of octahedra, *M[N(O)]$_6$* with *M = Mn, Fe, Co, Ni, Cu* or *Zn* and *M´[C(O)]$_6$* with *M´ = Fe$^{III}$, Fe$^{II}$,* and *Co$^{III}$,* arranged in an essentially cubic lattice and linked by cyanide ligands. Therefore, each metal has a coordination number of 6. In the above notation for the octrahedra, we included the possibility of *O* atoms instead of *C* or *N* since the oxygen atom of the water molecules may occupy those regular lattice positions in order to complete the octahedra. In $Fm\bar{3}m$ and $F\bar{4}3m$ type PBAs, the *O* occupancy of such regular positions occurs essentially randomly at defect sites; however, for $Pm\bar{3}m$ type PBAs, the octahedra of the divalent metal ion is formed by 3 *N* and 3 *O* atoms [20]. For some of our PBAs, the room temperature structural parameters had been published previously [21-33], and our parameters are found to be in good agreement with those data.

*2.2.2 X-ray fluorescence spectroscopy (XRF)*

X-ray fluorescence (XRF) measurements were performed using PANalytical's MiniPal QC energy dispersive X-ray fluorescence (EDXRF) benchtop spectrometer with rhodium tube and a silicon drift detector to determine the elemental composition (including potential potassium content) for each of our samples. Measurements were performed at room temperature on pressed powders using an X-ray tube operating voltage of 14 kV and a current of 150 µA. All of our single-phase PBAs exhibit compositions that equal the nominal compositions within error bars. XRF measurements indicate that the amount of potassium is negligible (less than 5%) for all of our PBAs.

*2.2.3 Fourier-transform infrared (FTIR) spectroscopy*

FTIR spectroscopy was used mainly to establish the presence of cyanide bonds and water molecules in each of our single phase PBAs. For all samples, FTIR spectra were collected in the mid-IR range 400–4000 cm$^{-1}$ using a Thermo Nicolet NEXUS 670 FT-IR spectrometer at ambient temperature and pressure. Figure 2 shows a typical FTIR spectra for $Zn_3[Fe(CN)_6]_2.14H_2O$, which confirms the presence of lattice water at 3650–3350 cm$^{-1}$ (antisymmetric and symmetric *O–H* stretching modes) and at 1606 cm$^{-1}$ (*H–O–H* bending mode) and the cyanide bond at 2160 cm$^{-1}$ (*C≡N* stretching mode).

*2.2.4 Thermogravimetric analysis (TGA)*

TGA data were collected for all the samples using NETZSCH STA 449 C Jupiter instrument. All the measurements were performed under argon gas atmosphere with few milligrams of the sample loaded in $Al_2O_3$ crucible. The temperature range used for the measurements was 25 – 400º C with a heating rate of 2º/min. All of our PBAs contain water molecules, and the water content is known to vary depending on the actual synthesis process and conditions. There are two distinct types of water in PBAs: lattice water located on special lattice positions in the PBA framework or interstitial water at interstitial positions within the PBA framework. In fact, water molecules are essential; otherwise, the framework might collapse. In particular, while it is possible to remove interstitial water through annealing at moderate temperatures (i.e. dehydrated PBAs), the removal of lattice water leads to a collapse of the PBA framework because it neutralizes the charges arising from defect (such as unfilled) octahedras. The number of water molecules (lattice and interstitial) per formula unit in a particular PBA can be determined from the analysis of TGA data. As shown in Figure 3, a typical TGA curve for *$Cu_3[Co(CN)_6]_2 \cdot 16H_2O$* shows a two-step mass loss, where the first step corresponds to the removal of interstitial water and the second step corresponds to the removal of lattice water. The mass losses can be taken to compute the water content of a given material. Table 2 lists the chemical compositions together with the number of water molecules for each of our PBAs. For most of our PBAs, we find that the number of water molecules varies between 10 and 18 water molecules. For all of our PBAs, our TGA data provide evidence that most of the water content (70-90%) can be attributed to interstitial water molecules, although there is a slight variation of the ratio of interstitial and lattice water for the different families of compounds. Moreover, the dehydration temperatures for different compounds of a given family show some variation. For example, the *hexacyanocobaltates (III)*, *$M^{II}_3[Co^{III}(CN)_6]_2 \cdot nH_2O$*, exhibit dehydration temperatures (determined from the onset of the second-step mass loss) of about 145º, 150º, 152º, 170º, 125º, and 90ºC for the *Mn, Fe, Co, Ni, Cu,* and Zn analogs, respectively. A similar range of dehydration temperatures is found for *hexacyanferrates(III)* and *hexacyanoferrates(II)*.

*2.3 X-ray Powder Diffraction at different Temperatures*

To monitor the changes of the lattice parameters for each of the PBAs, variable temperature XRD patterns were collected in *2θ* range 10-60° with a step size of 0.02° using the same Rigaku Ultima III X-ray diffractometer with monochromatic $CuK_α$ (*λ = 1.54 Å*) radiation. The scan speed used was in the range 0.3–0.5°/min. Powder samples were loaded on a copper plate, and measurements were done in vacuum while cooling using a cold stage that uses liquid nitrogen. For all of our PBAs, we collected X-ray diffraction patterns at 8 different temperatures between room temperature 25ºC (298 K) and -150ºC (123 K) with temperature steps of ~25ºC. All samples were mounted on the cold finger of the cryostat and rapidly cooled under vacuum in order to prevent water loss.

## 2.4 Structure Refinement

Lattice parameters of each compound at each temperature were refined using Rietveld method with the help of GSAS program [19]. The refinements were initiated in the space group $Fm\bar{3}m$ or $F\bar{4}3m$ with the available atomic positions in the ICSD database [21-33], whenever available. All of the PBAs studied here were previously reported in the ICSD database. However, for $Co_2[Fe(CN)_6].18H_2O$ and $Ni_2[Fe(CN)_6].18H_2O$, we did not find the reported $F\bar{4}3m$ space group, but instead we were able to fit the diffraction pattern with $Pm\bar{3}m$, similar to what had been listed for $Cu_2[Fe(CN)_6].16H_2O$. For all PBAs, lattice constant, thermal parameters, atom positions, peak profiles, and background parameters were refined. For a few samples, it was necessary to take into account a certain amount of preferred orientation in order to reproduce Bragg peak intensities accurately. This is not unexpected since larger powder particles containing several crystallites may suffer from some preferred orientation. In Figure 4, we show a typical example of calculated and observed intensities for the X-ray diffraction patterns for the case of $Cu_3[Co(CN)_6]_2.16H_2O$ at 300K. The refined lattice constants and positional parameters are in good agreement with previously published results [21-33] if those were available. In few cases, we obtained relatively large R-factors (up to 12%). Apart from complications due to preferred orientations, this may also be attributed to significant structural disorder in the sample. In any case, however, the refinement of the lattice parameters produces reasonably accurate values, even if atomic positions and thermal parameters are less reliable for these samples. Our refinements show no evidence for any structural phase transition over the studied temperature range.

## 3. Results and Discussions

In cubic crystals, it is possible to relate the bulk thermal expansion coefficient to changes in the unit cell volume. Using the refined lattice parameters at various temperatures, one can determine an average coefficient of thermal expansion

$$\alpha = \frac{\Delta a}{a_0 \Delta T},$$

where $\Delta a$ is the average change in the lattice parameters over a temperature range and $a_0$ is the lattice parameter at a reference temperature (in our case, the room temperature). As shown by the study on $Cu_3[Co(CN)_6]_2 \cdot 16H_2O$ in Figure 4 (inset), thermal expansion effects ascertain themselves by shifts of the peak positions in the diffraction patterns at different temperatures. In a constant wavelength experiment, shifts to the larger 2θ with decreasing temperature are indicative of positive thermal expansion and shifts to the smaller 2θ are consistent with negative thermal expansion.

Taking the temperature dependence of the lattice parameter, *a*, for each of our PBAs, the average value of the linear thermal expansion coefficients (between -150°C and room temperature) can be extracted from the slopes of the linear fits to the lattice parameter *vs. T* curves. For all compounds of the three PBA families, the temperature variation of the lattice parameter and the least-squares linear fits are shown in Figures 5-7. The average linear thermal expansion coefficients and its (statistical) errors are determined by linear fits to the lattice-parameter values at the 8 different temperatures. The computed values are listed in Table 3. We would like to point out that the cubic symmetry gives rise to only a few diffraction peaks and that the least-squares Rietveld refinement is known to greatly underestimate the error bars of the lattice parameters (least-squares errors occur only in the third or fourth digit of the lattice parameters). In addition, there is the possibility of systematic errors and thus the actual values of the lattice parameters may be accurate only to within a few percent. An alternative way of determining a more realistic error for the peak positions (and thus the lattice parameters) was described by Wilson [34] using parabolic fits to individual diffraction peaks. We used Wilson's approach and Gaussian fits to estimate the errors in the determination of the lattice parameters, and we find that the typical error in the X-ray diffraction data for our PBAs amounts to about 0.02 Å. This is the error that we used in Figs 5-7. In any case, since all diffraction studies are done in the same experimental set up, we believe that our experiments will correctly capture the general trend in thermal expansion behavior and it can provide a quantitative estimate of the thermal expansion coefficient over the measured temperature range.

It is known that the coefficient of linear thermal expansion does not necessarily vary linearly with temperature. However, the intent of this paper is not to study the temperature dependence of the thermal expansion coefficient in our PBAs in detail, but to identify those compounds that exhibit NTE in the explored temperature range and, incidentally, to obtain some (quantitative) idea of the magnitude of the effect. As evident from Figures 5-7, the results indicate that some PBAs exhibit NTE behavior, while others exhibit positive thermal expansion. Below, we summarize and discuss the thermal expansion behavior for the different first row *hexacyanometallates* studied here.

### 3.1 Hexacyanocobaltates(III)

As can be seen in Figure 5, for all of our *hexacyanocobaltates(III)*, the low temperature lattice parameters, $a$, are larger than their respective room temperature values and the temperature variation can be considered approximately linear given the expected uncertainty in its determination. Therefore, we find that all *hexacyanocobaltates(III)* exhibit NTE behavior, and linear fits over the whole temperature range result in large average negative linear thermal expansion coefficients ranging from $-39.7 \times 10^{-6}$ $K^{-1}$ for *$Co_3[Co(CN)_6]_2.12H_2O$* to about $-19.6 \times 10^{-6}$ $K^{-1}$ for *$Fe_3[Co(CN)_6]_2.14H_2O$* (see Table 3). Such values are larger than, or comparable to, the reported thermal expansion coefficient of $-20.4 \times 10^{-6}$ $K^{-1}$ for *$Cd(CN)_2$*, the largest isotropic coefficient reported to date [35]. For some of the *hexacyanocobaltates(III)*, the temperature variation of the lattice parameters deviate from linearity at the lowest temperatures, indicating a transition to possible positive thermal-expansion behavior at lower temperatures. For example, neglecting the values at lowest temperatures, the thermal-expansion coefficients amount to about $-48.0 \times 10^{-6}$ $K^{-1}$ for *$Mn_3[Co(CN)_6]_2.12H_2O$*. Similarly, large NTE coefficients can be obtained for some of the other *hexacyanocobaltates(III)* and those are listed in Table 3 as well.

Our results indicate a significant compositional dependence of NTE behavior observed in this family of PBAs. The magnitude of the NTE in the series varies in a wide range with the divalent $M^{II}$ cation type in the following order *Mn > Fe > Co > Ni > Cu < Zn* as demonstrated in Figure 8 (lower panel). This trend in the magnitude of the NTE correlates directly with the trend in the room temperature lattice parameter and inversely with Irving-Williams series [36] for metal-complex stability (Figure 8, upper panel). Moreover, the variation in the magnitude of NTE and the lattice parameters also tend to correlate with the ionic radii [37] of the individual $M^{II}$ cations. These correlations are depicted graphically in Figure 8(upper and lower panel). Considering the crystal field and the ionic radii, the Irving-Williams series indicates that the strength of the binding interaction between the divalent transition metal cations (for high-spin metal ions) and the ligand vary as follows: *Mn < Fe < Co < Ni < Cu > Zn*. This is opposite to the order displayed by the magnitude of NTE and room temperature lattice parameters for the PBAs with different divalent metal cations. Such correlations implies that the strength of the $M^{II}$–N≡C ligand binding interaction plays an important role for NTE behavior observed in this PBA family. The observed correlations also indicate that, when the strength of the binding interaction between the divalent cation and the cyanide ligand is reduced (weaker $M^{II}$–N bond), the structure becomes more flexible, thus favoring stronger NTE behavior. Such correlations are similar to what has been reported by Chapman *et al.* [15] for the PBAs of type *MPt(CN)$_6$* with *M = Mn, Fe, Co, Ni, Cu, Zn, Cd*.

## 3.2 *Hexacyanoferrates(III)* and *Hexacyanoferrates(II)*

Unlike *hexacyanocobaltates(III)*, neither *hexacyanoferrates(III)* nor *hexacyanoferrates(II)* show NTE behavior throughout the respective series. In fact, most of those compounds exhibit a positive slope (i.e. PTE behavior) over the studied temperature range.

In the case of *hexacyanoferrates(III)*, compounds with *M = Mn, Co,* and *Ni* show more common positive thermal expansion behavior, while clear negative thermal expansion is found for *M = Cu* or *Zn* (Figure 6). The thermal expansion behavior for *M = Fe* is non-monotonic and there is also a significant amount of scatter in the variation of lattice parameters over the studied temperature range. The peak profiles were found to be severely broadened, likely due to particle-size peak broadening due to the formation of nanometer-sized particles during the synthesis of this particular compound. In other words, the refinement of the lattice parameters for this particular compound is less reliable, and the observed non-monotonic thermal expansion behavior may be due to larger uncertainty in the fitting procedure. Nevertheless, using the generous error bars of 0.02 (see discussion above), it is possible to obtain an average thermal expansion coefficient over the studied temperature range for all of the *hexacyanoferrates(III)*. The sign of the thermal expansion coefficient for *M = Fe* and *Co* is not completely certain, given the relatively large errors. The linear fit thermal expansion coefficients range from negative values of -39.6 x $10^{-6}$ K$^{-1}$ for *Zn$_3$[Fe(CN)$_6$]$_2$.14H$_2$O* to positive values of about 47.8 x $10^{-6}$ K$^{-1}$ for *Mn$_3$[Fe(CN)$_6$]$_2$.14H$_2$O* (see Table 3). The occurrence of NTE and PTE in *hexacyanoferrates(III)* indicates a possibility to create a zero-expansion PBA material by virtue of mixing of different metal ions at the $M^{II}$ cation site.

The results for the divalent *hexacyanoferrates(II)* with $M$ = *Mn, Co, Ni, Cu* and *Zn* are shown in Figure 7. Unlike the trivalent *hexcyanometallates* discussed above, all of these compounds show a more common positive thermal expansion behavior over the whole temperature range. For these compounds, the average coefficients range from 19.5 x $10^{-6}$ $K^{-1}$ for *$Co_2[Fe(CN)_6]$.$18H_2O$* to 43.1 x $10^{-6}$ $K^{-1}$ for *$Zn_2[Fe(CN)_6]$.$5H_2O$* (see Table 3). For the two non-cubic compounds in this family of PBAs, namely *$Mn_2[Fe(CN)_6]$.$9H_2O$* to *$Zn_2[Fe(CN)_6]$.$5H_2O$*, we determined average thermal expansion coefficients by averaging the temperature variation of the different lattice parameters.

Attempts to find some general Irving-Williams-type correlations of the thermal expansion coefficients with the values of the lattice parameters fail for both, *hexacyanoferrates(III)* and *hexacyanoferrates(II)*. In particular, we do not find any such correlations for the PBAs with PTE behavior, while *hexacyanoferrates(III)* that exhibit NTE still follow such relationship, as will be discussed below. As can be seen in Figure 9a, the thermal expansion coefficient for *hexacyanoferrates(III)* is positive and decreases almost linearly on going from *Mn* to *Ni* (ignoring the less reliable *Fe* compound) and it crosses over to negative values for *Cu* and *Zn*. As shown in Figure 9b, all of the *hexacyanoferrates(II)* exhibit almost metal-independent PTE, with the exception of *Zn*, which is non-cubic.

## 4. Conclusions

We studied the structural properties and the thermal expansion behavior of 17 different first-row (*3d*) transition-metal *hexacyanometallates* (or Prussian Blue Analogs). With the exceptions *$Mn_2[Fe(CN)_6]$.$9H_2O$* and *$Zn_2[Fe(CN)_6]$.$5H_2O$*, all other studied PBAs were found to crystallize as cubic structures. In the temperature range between -150°C and room temperature, negative thermal expansion coefficients were found in all of the trivalent *hexacyanocobaltates(III)* ((with $M$ = *Mn, Fe, Co, Ni, Cu* or *Zn*) as well as for some of the trivalent *hexacyanoferrates(III)* (with $M$ = *Fe, Cu* or *Zn*). All other studied compounds, including all of the divalent *hexacyanoferrates(II)*, show positive thermal expansion behavior.

As discussed above, all PBAs that exhibit NTE, i.e. all of the *hexacyanocobaltates(III)* and a couple of the *hexacyanoferrates(III)*, follow Irving-Williams-type correlations of the thermal-expansion coefficients with the lattice parameters. This is consistent with the universal relationship reported by Matsuda *et al.* [18]. As shown in Figure 10, our data indeed imply a similar universal scaling of the linear thermal expansion coefficients with the room-temperature lattice parameters for all of our PBAs with NTE. Matsuda *et al.* [18] proposed that the mechanism(s) responsible for NTE behavior can be attributed to some lattice instability. An alternative mechanism has been proposed by Chapman *et al.* [15], who proposed transverse vibrational motions to be ultimately responsible for NTE behavior. The exact nature of the underlying mechanisms for NTE behavior in many of our PBAs requires further investigation. An even more pressing question may be what drives the other (isostructural) PBAs to exhibit PTE behavior, as they do not show any Irving-Williams type correlations.

The main focus of this paper was to categorize, which of the studied PBAs exhibit NTE behavior. It lays the foundation for further investigations as to what mechanism(s) may be responsible for the thermal-expansion behavior in such materials. Not all of our PBAs exhibit NTE behavior and the thermal expansion coefficients span over a large range of values. Therefore, it seems evident that neither simple structural properties nor the full water content are the sole driving mechanisms for NTE. At best, the presence of water molecules may have only subtle effects on the shape of the lattice parameter *vs. T* curves, and we plan to investigate such effects in some future work. On the other hand, it seems that the valence of the metal ions may contribute given the fact that we observe NTE behavior only for the trivalent PBAs, although not all of those show NTE. Moreover, the actual valence state for each of our PBAs is only implied and has not yet been measured experimentally. XANES and voltammetric studies are underway to establish actual valences for each of our compounds. However, most likely, thermal expansion in our PBAs is driven by dynamical effects, similar to what has been reported for other related PBAs [15]. Combining inelastic neutron scattering and theoretical studies in the near future, we plan to shed some light into possible correlations between thermal expansion and the occurrence and nature of soft vibrational modes of our PBAs.

**Acknowledgement**

This research work has been supported by Department of Energy's (DOE) Office of Basic Energy Sciences and has made use of Manuel Lujan, Jr. Neutron Scattering Center at Los Alamos National Laboratory which is funded by DOE Office of Basic Energy Sciences. Los Alamos National Laboratory is operated by Los Alamos National Security, LLC, under DOE Contract DE-AC52-06NA25396. JS acknowledges partial support provided by Shengnian Luo at LANL (DOE grant number: 20110585ER).

**Figure Captions**

**Figure 1.** Schematic representations of Prussian Blue Analog frameworks showing alternating $M[N(O)]_6$ (light color) and $M'[C(O)]_6$ (dark color) octahedra: (a) with space group $Fm\bar{3}m$ (no. 225), (b) with space group $F\bar{4}3m$ (no. 216), and (c) with space group $Pm\bar{3}m$ (no. 221). Note that $F\bar{4}3m$ has an additional metal atom in the center of the cube with respect to $Fm\bar{3}m$. Possible $O$ positions (from the lattice water) at defect sites in $Fm\bar{3}m$ is indicated by darker semispheres at octrahedra corners, while possible $O$ positions (separate symbols) in $M'[C(O)]_6$ octahedra for $Pm\bar{3}m$ would result in slightly distorted octrahedra. For clarity, possible $O$ positions due to interstitial water molecules are not included in the drawings.

**Figure 2.** Fourier-transform infrared (FTIR) spectrum for $Zn_3[[Fe(CN)_6]_2.14H_2O$ showing vibrational modes that evidence the presence of water and cyanide ligands in the sample.

**Figure 3.** Exemplary thermogravimetric analysis (TGA) study for $Cu^{II}_3[Co^{III}(CN)_6]_2.16H_2O$ showing the two-step process of removing interstitial and lattice water molecules upon heating. In the inset, the possible area for interstitial water molecules in the $Fm\bar{3}m$ structure is indicated by the large sphere in the center.

**Figure 4.** Exemplary X-ray diffraction pattern (symbols) and Rietveld-refinement fit (solid line) for $Cu_3[Co(CN)_6]_2.16H_2O$. The line at the bottom represents the difference profile between the experimental data and the fit (off-set from zero for clarity). The inset shows X-ray patterns of the same compound at low temperature (black line) and room temperature (grey line) indicating the shifts of the peaks to the lower 2θ values at low temperatures.

**Figure 5.** Temperature dependence of the optimized lattice parameter *a* as obtained from Rietveld refinements of the XRD data at different temperatures for $M^{II}_3[Co^{III}(CN)_6]_2 \cdot nH_2O$ (*M* = *Mn, Fe, Co, Ni, Cu* and *Zn*). The number of water molecules (*n*) for the different compounds can be found in Table 2. The lines represent a linear fit of the data over the whole temperature range, and they were taken to compute the linear thermal expansion coefficients listed in Table 3. Error bars for a few selected lattice parameters are shown.

**Figure 6.** Temperature dependence of the optimized lattice parameter *a* as obtained from Rietveld refinements of the XRD data at different temperatures for $M^{II}_3[Fe^{III}(CN)_6]_2 \cdot nH_2O$ (*M* = *Mn, Fe, Co, Ni, Cu* and *Zn*). The number of water molecules (*n*) for the different compounds can be found in Table 2. The lines represent a linear fit of the data over the whole temperature range, and they were taken to compute the linear thermal expansion coefficients listed in Table 3. Error bars for a few selected lattice parameters are shown.

**Figure 7.** Temperature dependence of the optimized lattice parameter *a* as obtained from Rietveld refinements of the XRD data at different temperatures for $M^{II}_2[Fe^{III}(CN)_6] \cdot nH_2O$ (*M* = *Mn, Co, Ni, Cu* and *Zn*). The number of water molecules (*n*) for the different compounds can be found in Table 2. The lines represent a linear fit of the data over the whole temperature range, and they were taken to compute the linear thermal expansion coefficients listed in Table 3. Note the different *y*-axis for different compounds. Error bars for a few selected lattice parameters are shown.

**Figure 8.** Trends and/or correlations between Irving-Williams series, ionic radii of the $M^{II}$ cations, CTEs and lattice parameters of the studied PB analogs: (upper panel) variation of log stability constant, i.e. Irving-Williams series (■) and ionic radii (○) with $M^{II}$ cations in the order of increasing atomic number (after refs 36 and 37); (lower panel) compositional dependence of the coefficient of thermal expansion (CTE or $\alpha$) (♦) and room-temperature lattice parameter (●) for the *hexacyanocobaltates(III)* showing correlations with the Irving-Williams series.

**Figure 9**. Coefficient of thermal expansion (CTE or $\alpha$) for (a) the compounds of *hexacyanoferrates(III)* and (b) the compounds of *hexacyanoferrates(II)* as a function of $M^{II}$ cations.

**Figure 10**. Variation of coefficient of thermal expansion (CTE or $\alpha$) with the room-temperature lattice parameters for all the PBAs showing NTE behavior: *hexacyanocobaltates(III)* (♦) and *hexacyanoferrates(III)* (●).

**Tables and captions**

**Table 1.** Space groups, atom sites and general atom positions for Prussian Blue Analogs with general formulas $M^{II}_3[(M´)^{III}(CN)_6]_2 \cdot nH_2O$ and $M^{II}_2[Fe^{II}(CN)_6] \cdot nH_2O$ with M = Mn, Fe, Co, Ni, Cu or Zn and M´ = Co or Fe. Possible positions for H and O atoms of water molecules are not included in the table. In general, there will be some O from the lattice water molecules occupying some of the regular C or N lattice positions. The refined parameters for general x parameters for the C and N atoms for some of the PBAs are provided in Table 2.

| Atoms | Site | x | y | z |
|---|---|---|---|---|
| *Space group: $Fm\bar{3}m$ (cubic, no. 225)* | | | | |
| A | 4a | 0 | 0 | 0 |
| M | 4b | ½ | ½ | ½ |
| C | 24e | 0 | 0 | $x_C$ |
| N | 24e | 0 | 0 | $x_N$ |
| *Space group: $F\bar{4}3m$ (cubic, no. 216)* | | | | |
| A | 4a | 0 | 0 | 0 |
| M1 | 4b | ½ | ½ | ½ |
| M2 | 4c | ¼ | ¼ | ¼ |
| C | 24f | $x_C$ | 0 | 0 |
| N | 24f | $x_N$ | 0 | 0 |
| *Space group: $Pm\bar{3}m$ (cubic, no. 221)* | | | | |
| M1 | 1a | 0 | 0 | 0 |
| M2 | 3c | 0 | ½ | ½ |
| Fe1 | 3d | 0 | 0 | ½ |
| Fe2 | 1b | ½ | ½ | ½ |
| C1 | 6e | 0 | 0 | $x_{C1}$ |
| N2 | 6e | 0 | 0 | $x_{N1}$ |
| C2 | 12h | 0 | ½ | $x_{C2}$ |
| N2 | 12h | 0 | ½ | $x_{N2}$ |
| C3 | 6f | ½ | ½ | $x_{C3}$ |
| N3 | 6f | ½ | ½ | $x_{N3}$ |
| *Space group: $P\bar{3}$ (trigonal, no. 147)[a]* | | | | |
| Zn | 2d | 1/3 | 2/3 | 0.3988 |
| Fe | 1a | 0 | 0 | 0 |
| C | 6g | 0.2162(90) | 0.2622(49) | 0.2503(44) |
| N | 6g | 0.2436 | 0.4515 | 0.1659 |
| *Space group: $P2_1/n$ (monoclinic, no. 14)[b]* | | | | |
| Fe | 2a | ½ | ½ | ½ |
| Mn | 4e | 0.6295 | 0.0855 | 0.4317 |
| C1 | 4e | 0.4431 | 0.4480 | -0.0664 |
| C2 | 4e | 0.7879 | 0.5471 | 0.4277 |
| C3 | 4e | 0.3324 | 0.3937 | 0.3160 |
| N1 | 4e | 0.8468 | 0.2107 | 0.4721 |
| N2 | 4e | 0.6884 | 0.5746 | 0.2886 |
| N3 | 4e | 0.3315 | 0.3553 | 0.2439 |

[a] only $Zn_2[Fe(CN)_6] \cdot 5H_2O$ crystallizes in this structure
[b] only $Mn_2[Fe(CN)_6] \cdot 9H_2O$ crystallizes in this structure

**Table 2.** Compounds, space group, room-temperature lattice parameter and room-temperature atomic positions of the *C* and *N* atoms. Note that some of the *C* and *N* positions may be occupied by *O* from the lattice water; either at defect sites or in a regularly ordered fashion such as for $Pm\bar{3}m$ type PBAs (see text).

| PB analogs | Space group | $a$ [Å] | $x_C$ | $x_N$ |
|---|---|---|---|---|
| *Hexacyanocobaltates(III)* | | | | |
| $Mn_3[Co(CN)_6]_2 \cdot 12H_2O$ | $F\bar{4}3m$ | 10.2876(5) | 0.1915 | 0.2996 |
| $Fe_3[Co(CN)_6]_2 \cdot 14H_2O$ | $F\bar{4}3m$ | 10.1944(9) | 0.1980 | 0.2996 |
| $Co_3[Co(CN)_6]_2 \cdot 12H_2O$ | $F\bar{4}3m$ | 10.0748(22) | 0.1928 | 0.3060 |
| $Ni_3[Co(CN)_6]_2 \cdot 16H_2O$ | $F\bar{4}3m$ | 10.0092(14) | 0.1961 | 0.3033 |
| $Cu_3[Co(CN)_6]_2 \cdot 16H_2O$ | $Fm\bar{3}m$ | 9.9624(8) | 0.1777 | 0.2955 |
| $Zn_3[Co(CN)_6]_2 \cdot 9H_2O$ | $F\bar{4}3m$ | 10.0976(10) | 0.1835 | 0.3012 |
| *Hexacyanoferrates(III)* | | | | |
| $Mn_3[Fe(CN)_6]_2 \cdot 14H_2O$ | $F\bar{4}3m$ | 10.4539(4) | 0.1809 | 0.2860 |
| $Fe_3[Fe(CN)_6]_2 \cdot 14H_2O$ | $Fm\bar{3}m$ | 10.0747(56) | 0.3257 | 0.2149 |
| $Co_3[Fe(CN)_6]_2 \cdot 17H_2O$ | $F\bar{4}3m$ | 10.0884(19) | 0.0904 | 0.2700 |
| $Ni_3[Fe(CN)_6]_2 \cdot 14H_2O$ | $F\bar{4}3m$ | 10.1962(14) | 0.2010 | 0.2717 |
| $Cu_3[Fe(CN)_6]_2 \cdot 18H_2O$ | $Fm\bar{3}m$ | 10.0316(6) | 0.2079 | 0.3208 |
| $Zn_3[Fe(CN)_6]_2 \cdot 14H_2O$ | $Fm\bar{3}m$ | 10.1009(19) | 0.1759 | 0.3085 |
| *Hexacyanoferrates(II)* | | | | |
| $Mn_2[Fe(CN)_6] \cdot 9H_2O$ | $P2_1/n$ [a] | 9.9188(50) [c] | see table 1 | |
| $Co_2[Fe(CN)_6] \cdot 18H_2O$ | $Pm\bar{3}m$ | 10.2630(11) | *1:* 0.4212 | 0.2015 |
| | | | *2:* 0.1768 | 0.3099 |
| | | | *3:* 0.2863 | 0.1819 |
| $Ni_2[Fe(CN)_6] \cdot 18H_2O$ | $Pm\bar{3}m$ | 10.0701(44) | *1:* 0.2630 | 0.2304 |
| | | | *2:* 0.1242 | 0.3158 |
| | | | *3:* 0.3141 | 0.2018 |
| $Cu_2[Fe(CN)_6] \cdot 16H_2O$ | $Pm\bar{3}m$ | 9.9556(12) | *1:* 0.2209 | 0.1630 |
| | | | *2:* 0.2324 | 0.3469 |
| | | | *3:* 0.3034 | 0.1846 |
| $Zn_2[Fe(CN)_6] \cdot 5H_2O$ | $P\bar{3}$ [b] | 6.9625(17) [c] | see table 1 | |

[a] monoclinic structure with $a = 9.1407(37)$ Å, $b = 12.3725(88)$ Å, $c = 8.2042(39)$ Å, $\beta = 102.99(3)°$
[b] hexagonal (or trigonal) structure. $a = b = 7.5564(13)$ Å, $c = 5.7748(26)$ Å; $\alpha = \beta = 90°$, $\gamma = 120°$
[c] not cubic; reported parameter is the average of the $a$, $b$, and $c$ lattice parameters

**Table 3.** Coefficients of thermal expansion ($\alpha$, in units of $10^{-6}$ K$^{-1}$) for the PB analogs as determined from linear fits to the Rietveld-refined lattice parameters as a function of temperature in the whole studied T-range (298–123 K).

| M  | M$_3$[Co(CN)$_6$]$_2$ | M$_3$[Fe(CN)$_6$]$_2$ | M$_2$[Fe(CN)$_6$] |
|----|----|----|----|
| Mn | -29.2(5.8)<br>-48.0(2.5) [a] | +47.8(3.4) | +20.2(5.0) [b] |
| Fe | -19.6(7.0)<br>-39.3(6.0) [a] | -9.9(12.9) | [c] |
| Co | -39.7(5.2)<br>-35.5(5.1) [a] | +7.9(5.1) | +19.5(10.7) |
| Ni | -30.0(4.4) | +5.9(1.9) | +19.9(3.0) |
| Cu | -20.0(1.2) | -19.9(0.6) | +20.1(9.7) |
| Zn | -29.7(2.8)<br>-33.7(2.6) [a] | -39.6(6.2) | +43.1(2.7) [b] |

[a] these values are determined from the least square linear fits excluding the data points below 175, 198, 148, and 148 K for *M = Mn, Fe, Co,* and *Zn* respectively.
[b] expansion coefficient was determined from average lattice-parameter variation
[c] compound is unstable at ambient conditions

**List of figures**

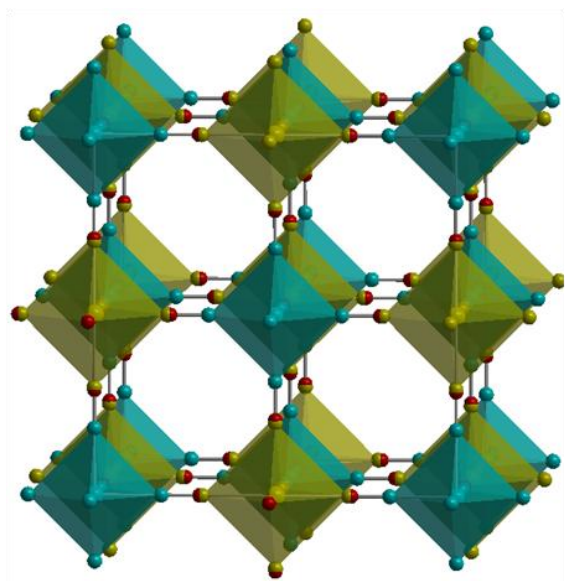

a) Space group: $F m\bar{3} m$

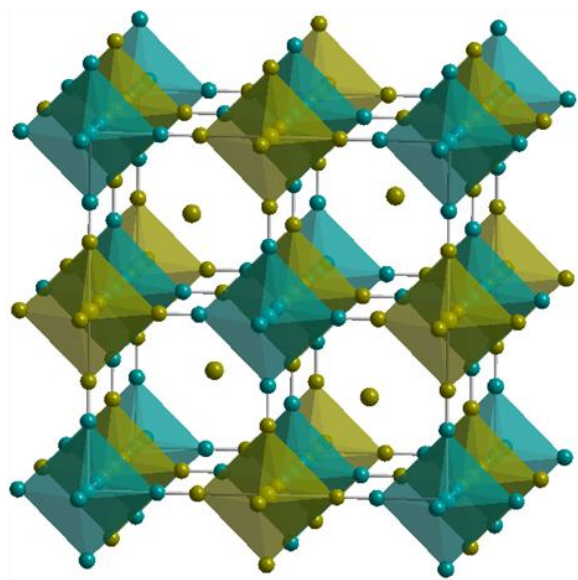

b) Space group: $F\bar{4}3m$

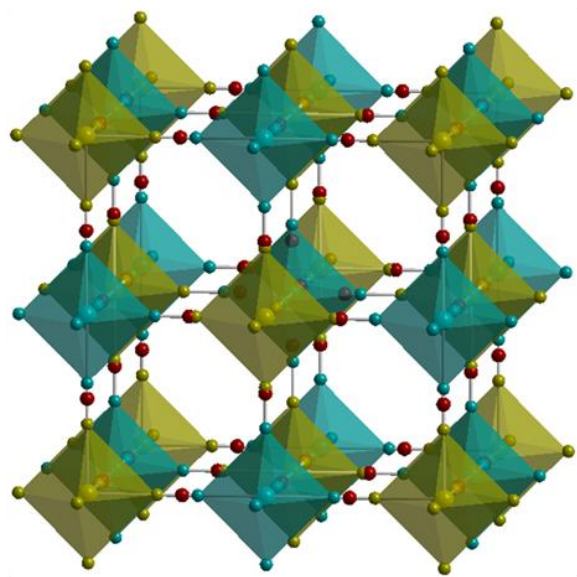

c) Space group: $P m\bar{3} m$

**Figure 1**

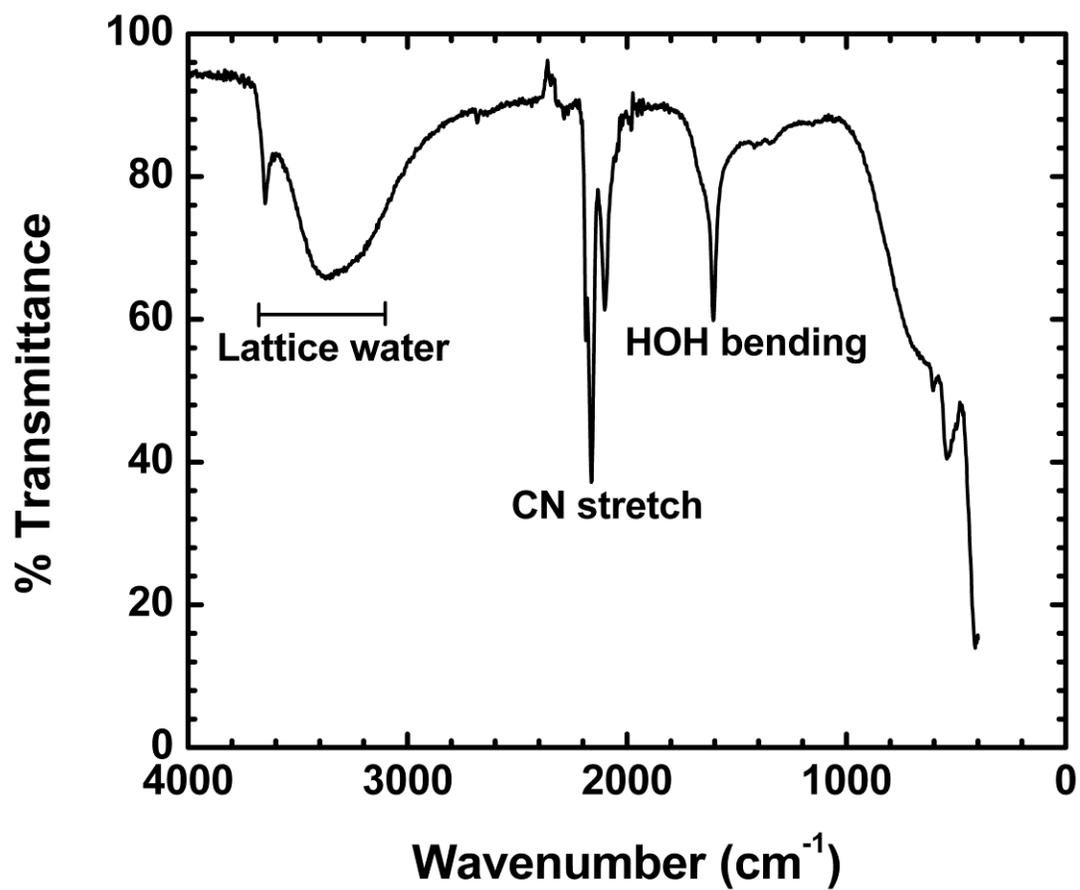

**Figure 2**

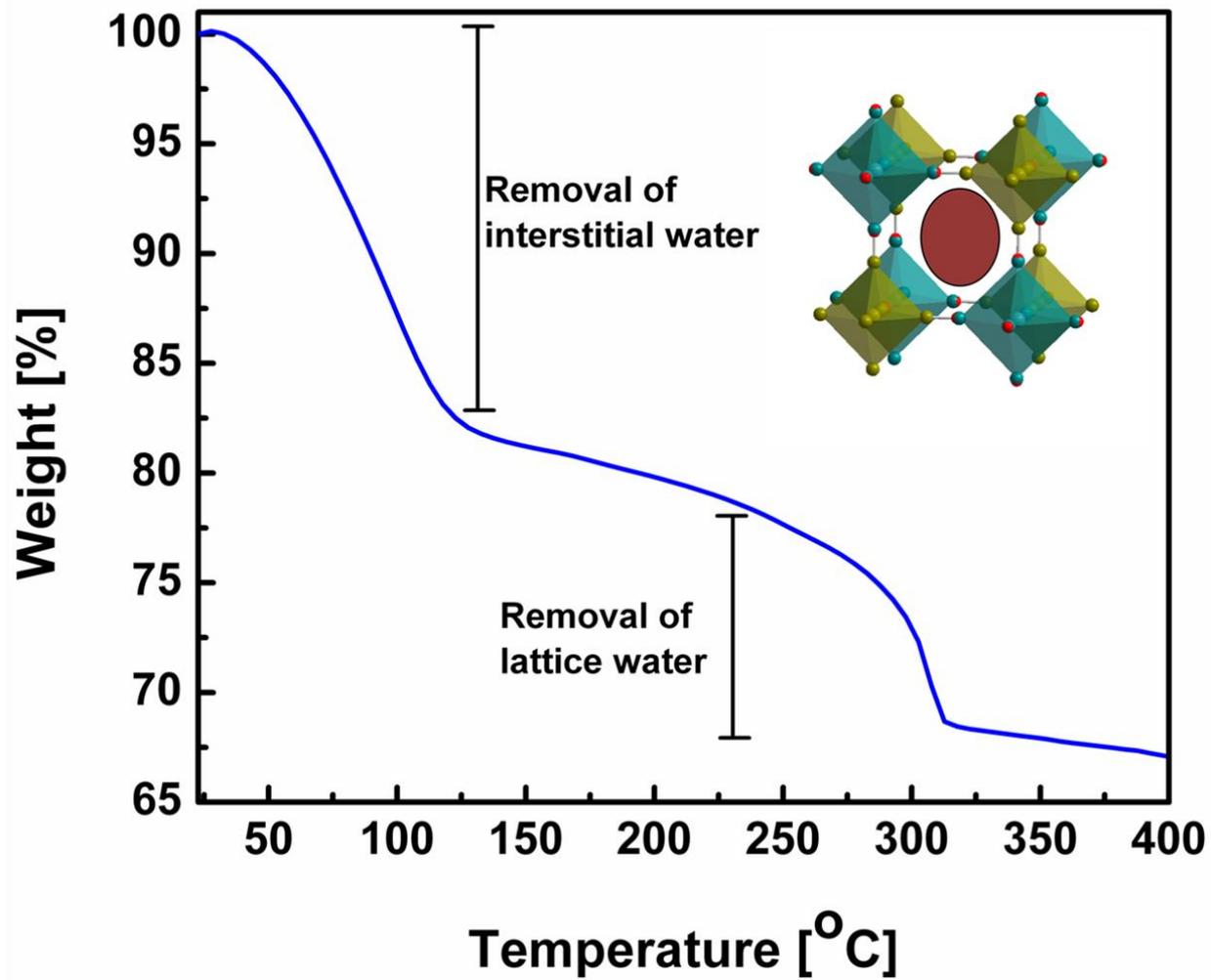

**Figure 3**

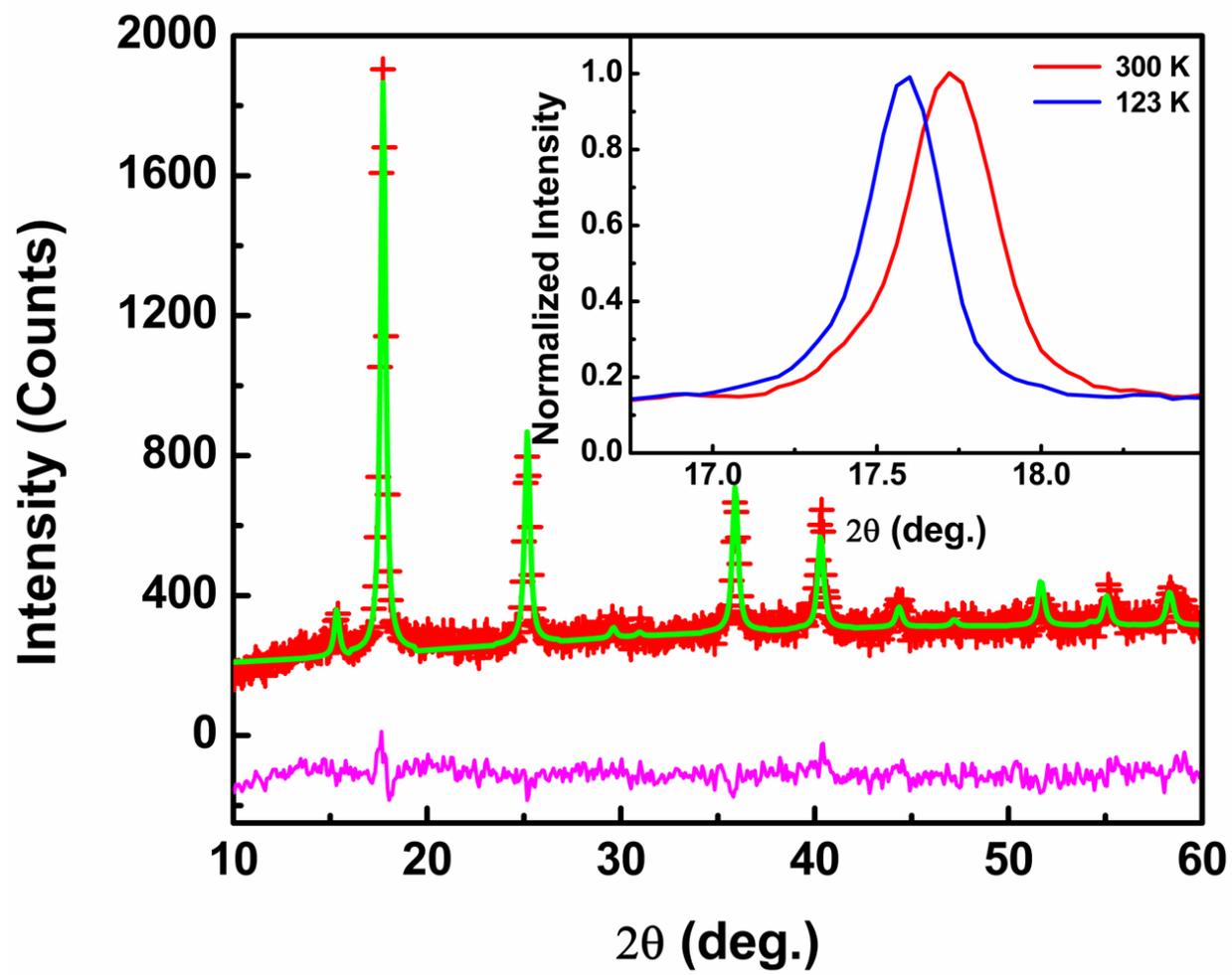

**Figure 4**

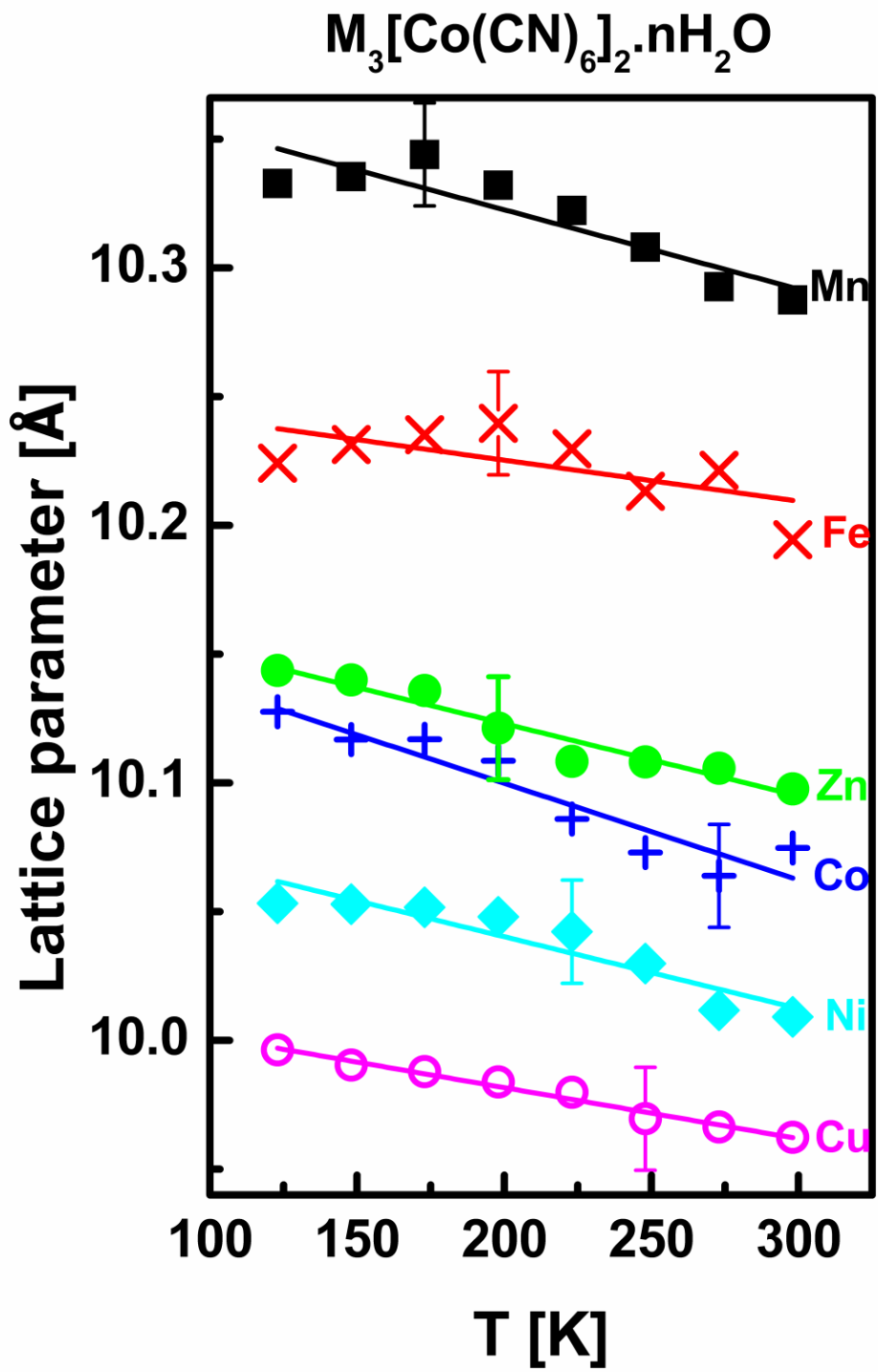

**Figure 5**

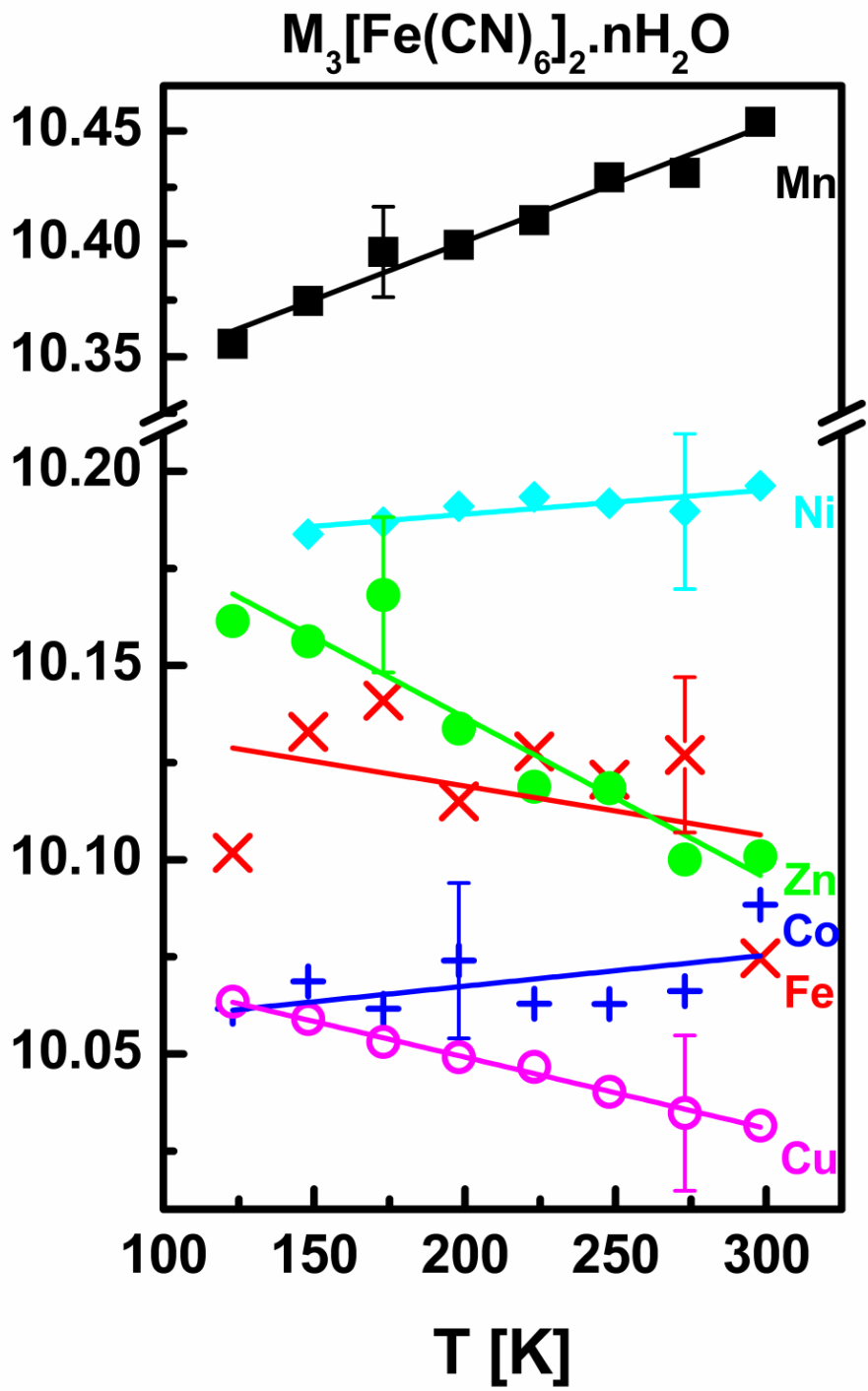

**Figure 6**

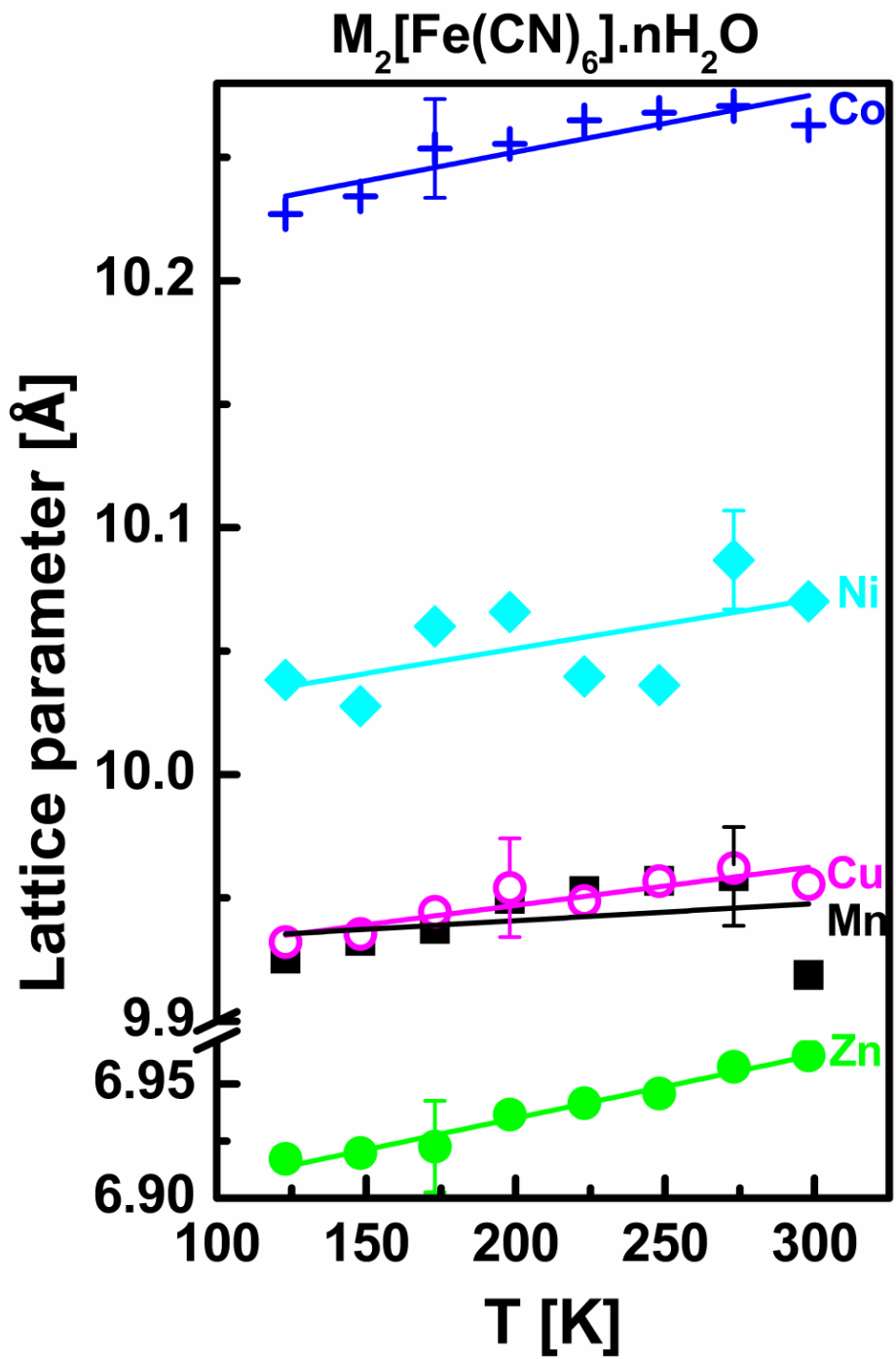

**Figure 7**

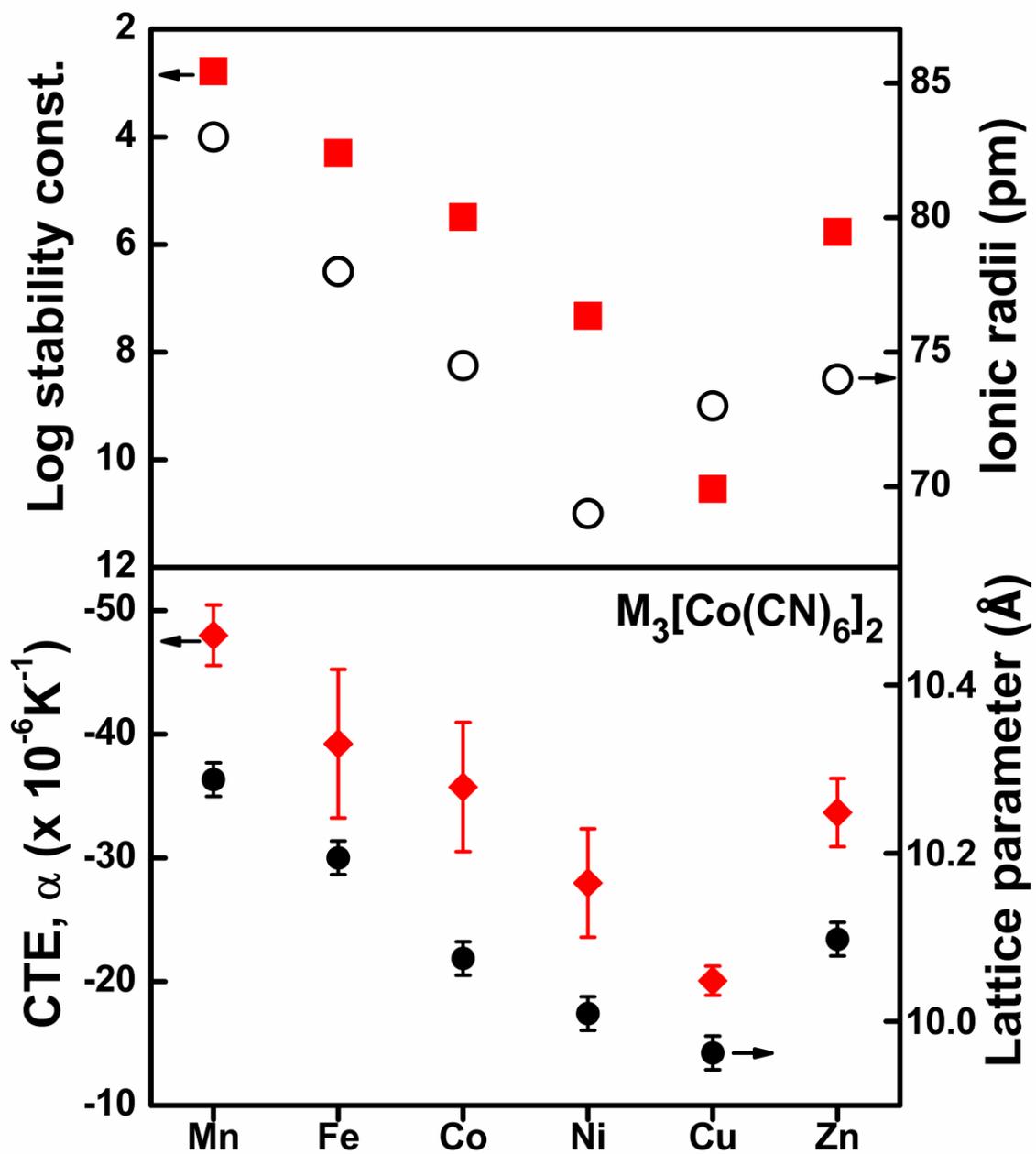

**Figure 8**

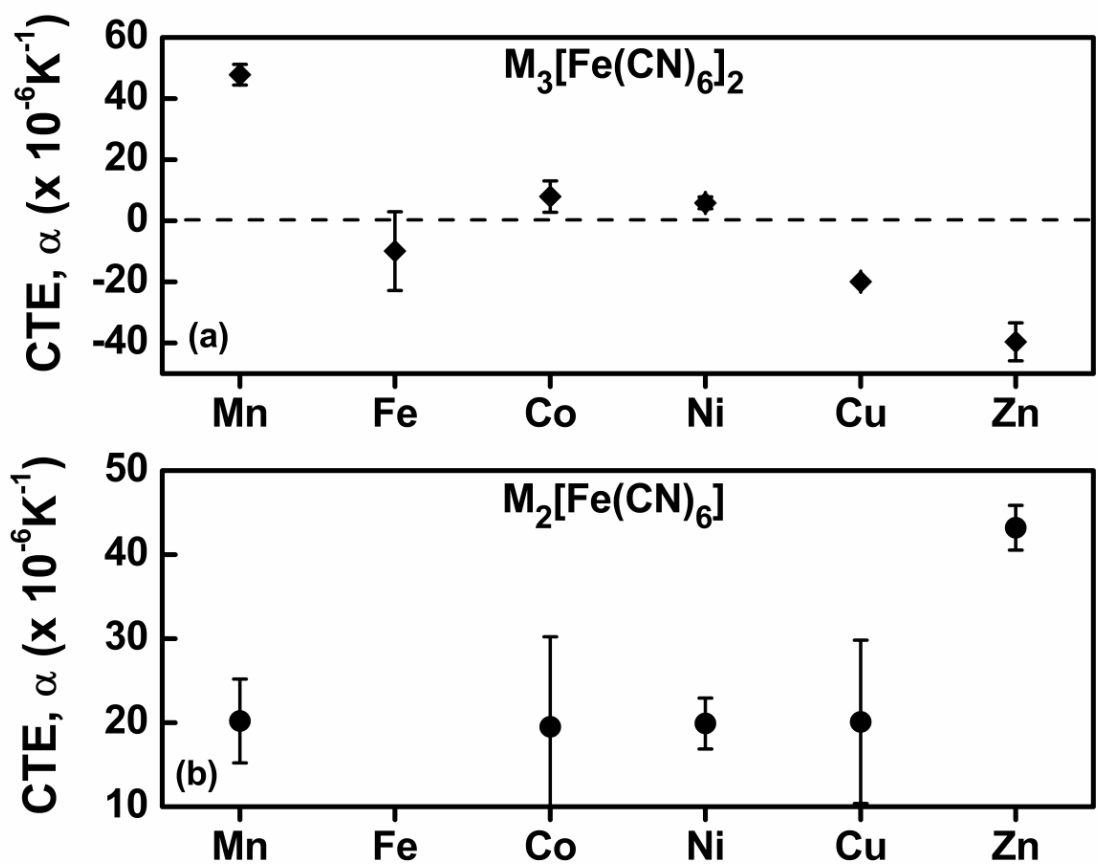

**Figure 9**

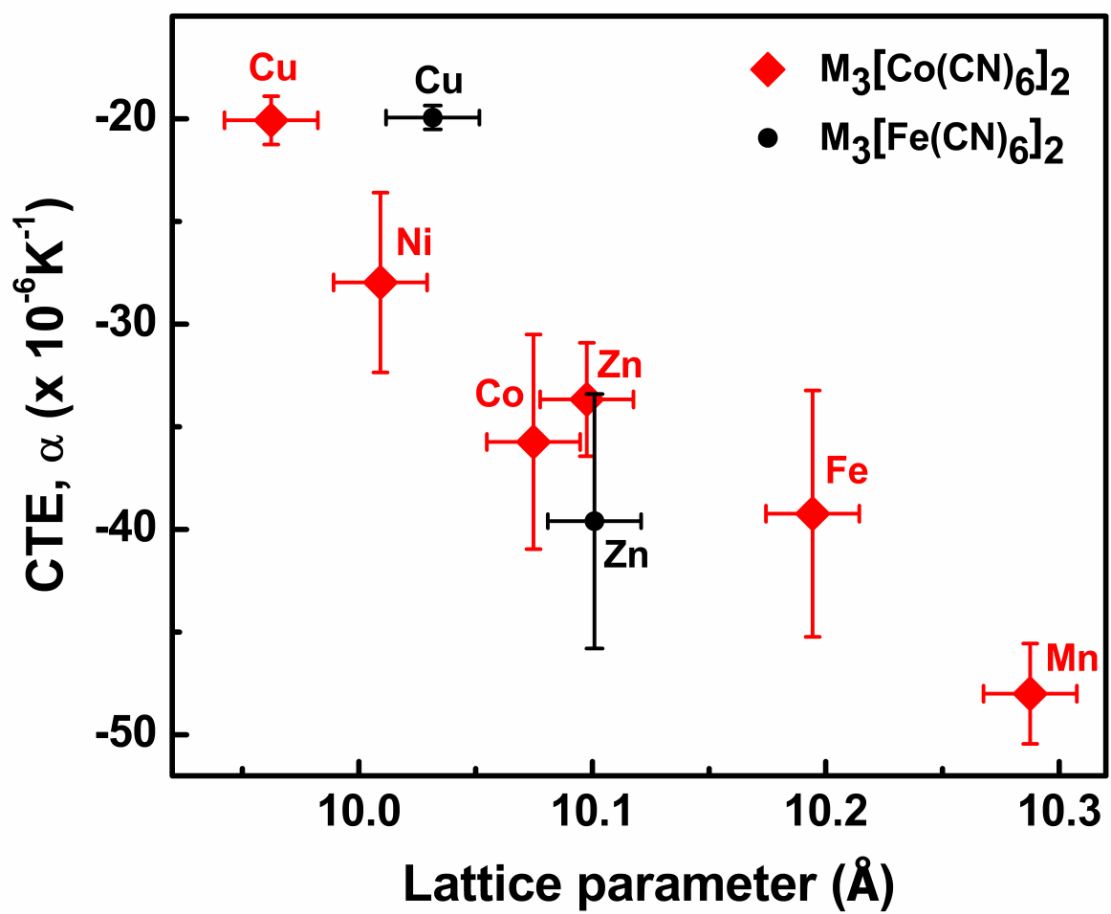

Figure 10